	\documentclass[12pt,preprint]{aastex}
	\shorttitle{HD 1383}
	\shortauthors{Boyajian et al.}
\begin{document}


\title{The B-Supergiant Components of the 
Double-Lined Binary HD~1383} 

\author{T. S. Boyajian, D. R. Gies\altaffilmark{1}, 
 M. E. Helsel\altaffilmark{2}, A. B. Kaye\altaffilmark{1}, \\
 M. V. McSwain\altaffilmark{1,3,4}, 
 R. L. Riddle\altaffilmark{1,5}, D. W. Wingert\altaffilmark{1}}
\affil{Center for High Angular Resolution Astronomy and \\
 Department of Physics and Astronomy,\\
 Georgia State University, P. O. Box 4106, Atlanta, GA 30302-4106; \\
 tabetha@chara.gsu.edu, gies@chara.gsu.edu, marian.helsel@furman.edu, anthony.kaye@itt.com,  
 mcswain@astro.yale.edu, riddle@astro.caltech.edu, wingert@chara.gsu.edu}

\altaffiltext{1}{Visiting Astronomer, Kitt Peak National Observatory,
National Optical Astronomy Observatory, operated by the Association
of Universities for Research in Astronomy, Inc., under contract with
the National Science Foundation.}
\altaffiltext{2}{Current Address: Department of Chemistry, 
Furman University, 3300 Poinsett Highway, Greenville, SC 29613}
\altaffiltext{3}{Current Address: Astronomy Department,
Yale University, New Haven, CT 06520-8101}
\altaffiltext{4}{NSF Astronomy and Astrophysics Postdoctoral Fellow}
\altaffiltext{5}{Current Address: Department of Astronomy, 
California Institute of Technology, 305 S.\ Hill Ave., MC 102-8,
Pasadena, CA 91125}

\def\hzo        {H$_2$O}
\def\kms    {\ifmmode{{\rm km~s}^{-1}}\else{km~s$^{-1}$}\fi}
\def\Mdot   {\ifmmode {\dot M} \else $\dot M$\fi}
\def\Mspy   {\ifmmode {M_{\odot} {\rm yr}^{-1}} \else $M_{\odot}$~yr$^{-1}$\fi}
\def\Msun   {$M_{\odot}$}
\def\mum     {\ifmmode{\mu{\rm m}}\else{$\mu{\rm m}$}\fi}
\def\Rstar  {$R_{\star}$}
%

\begin{abstract}

We present new results from a study of high quality, 
red spectra of the massive binary star system HD~1383 
(B0.5~Ib + B0.5~Ib).  We determined radial velocities 
and revised orbital elements ($P = 20.28184\pm 0.0002$~d) 
and made Doppler tomographic reconstructions of the 
component spectra.  A comparison of these with model
spectra from non-LTE, line blanketed atmospheres 
indicates that both stars have almost identical 
masses ($M_2/M_1 = 1.020\pm 0.014$), 
temperatures ($T_{\rm eff} = 28000 \pm 1000$~K), 
gravities ($\log g = 3.25 \pm 0.25$), and 
projected rotational velocities ($V\sin i \lesssim 30$ \kms). 
We investigate a number of constraints on the radii 
and masses of the stars based upon the absence of eclipses, 
surface gravity, stellar wind terminal velocity, and 
probable location in the Perseus spiral arm of the Galaxy, 
and these indicate a range in probable radius and mass
of $R/R_\odot = 14 - 20$ and $M/M_\odot = 16 - 35$, respectively.
These values are consistent with model evolutionary 
masses for single stars of this temperature and gravity. 
Both stars are much smaller than their respective Roche 
radii, so the system is probably in a pre-contact stage 
of evolution.  A fit of the system's spectral energy 
distribution yields a reddening of $E(B-V)=0.55\pm0.05$ and 
a ratio of total-to-selective extinction of $R=2.97\pm0.15$.
We find no evidence of H$\alpha$ emission from colliding 
stellar winds, which is probably the consequence of the 
low gas densities in the colliding winds zone. 

\end{abstract}



\setcounter{footnote}{5}

\section{Introduction}                              

The evolutionary paths of massive binaries depend critically 
on processes related to mass transfer and mass loss at the time 
when the initially more massive star begins to fill its Roche volume 
\citep*{lan04,pet05}.  On the one hand, mass from the donor 
star may end up entirely in a rejuvenated mass gainer, but 
on the other hand a high mass transfer rate may cause the 
gainer to swell and bring the system into a common 
envelope phase, in which most of the donor's mass is lost from 
the system entirely.  Evidence of both outcomes is found among 
post-Roche lobe overflow systems \citep{lan04,pet05}.  
We can better constrain the problem of the probable results 
of the interaction by studying binaries in an advanced 
evolutionary state just prior to Roche-filling.  Those 
binaries of nearly identical masses are particularly interesting
because we can usually observe the spectra of both components.  

Close pairs of nearly identical, evolved massive stars are 
quite rare.  The HD~1383 system is the only such system 
found in {\it Ninth Catalogue of Spectroscopic Binary 
Orbits} \citep{pou04} (with identical components of types 
B0.5~Ib + B0.5~Ib; \citealt{hil86}).
Another similar massive stellar system, HD~152248,  
was previously classified as a O7~I + O7~I 
\citep*{pen99}, although a more thorough analysis by 
\citet{san01} revised the classifications to 
O7.5~III + O7~III, deviating from the exact match of types found 
in HD~1383.  The HD~152248 system is eclipsing and the mass 
determinations made by \citet{pen99} and \citet{san01} 
indicate that both stars are undermassive (by factors 
of 2 and 1.4, respectively, in the two studies) relative to 
the predicted evolutionary masses for single stars. 
Here we focus our attention on HD~1383 
(BD$+60\degr25$, HIP~1466).  The system was identified as
a double-lined binary early on \citep{san38,sle56}, 
but the first orbital elements were determined later 
by \citet{hil86} who found that the system consists of 
two nearly identical stars with an orbital period of 20.3~d. 
\citet*{mor55} originally adopted the system as a spectral standard 
for the B1~II type, but \citet{hil86} revised 
the types slightly based upon detailed measurements of 
line equivalent widths of both components.   The system is 
located in the sky in the vicinity of the Cas~OB4 association  
\citep{hum78}, which is located in the Perseus arm of the Galaxy.  
The star has played an important role in studies of the 
interstellar medium in this direction \citep[see][]{car04}.  
Various distances have been estimated based upon 
the assumption that HD~1383 is a single, B1~II star.
\citet{wak98} found a distance of 1.7~kpc, which agrees
well with \citet{hum78} value of 1.68~kpc.  
\citet{hum78} noted that HD~1383 might not be a 
member of the Cas~OB4 association, because the latter has a
greater distance of 2.88~kpc.  However, the fact that HD~1383 
is a binary consisting of two stars with approximately
equal brightness implies that it farther away from us than 
these previous estimates indicate and that it is closer to 
Cas~OB4 (\citealt{hil86} offer a distance 3.0~kpc for
HD~1383).  

Here we present a study of a set of red spectra of HD~1383 
that we obtained (\S2) to re-analyze the orbital elements (\S3) and 
to search for evidence of H$\alpha$ emission related to colliding winds
\citep{tha97,san01}.  We discuss the physical parameters of the stars 
from an analysis of their individual spectra obtained 
from a Doppler tomography reconstruction (\S4).  
We then present a number of constraints on the radii and 
masses of the stars (\S5) that lead us to conclude that 
both stars have radii smaller than their Roche radii and
luminosities that are too small to power strong winds 
(explaining the lack of H$\alpha$ emission from colliding winds). 


\section{Observations and Radial Velocities}        

The optical spectra of HD~1383 were obtained with the Kitt Peak 
National Observatory 0.9~m coud\'{e} feed telescope during four observing 
runs between 1999 August and 1999 December.  Most of these spectra were 
obtained with the short collimator and grating RC181 
(316 grooves mm$^{-1}$ with a blaze wavelength of 7500 \AA ; made in first order 
with a GG495 filter to block higher orders), which yielded an average 
resolving power of $R=\lambda / \delta \lambda = 4000$ 
(see \citealt{gie02a} for details of each observing run). 
The last four spectra were made with grating B 
(632 grooves mm$^{-1}$ with a blaze wavelength of 6000 \AA ~in second order)
and these have a much higher resolving power, $R=\lambda / \delta \lambda = 21300$.  
Exposure times varied from 20 to 30 minutes.  These spectra 
all cover a common spectral range between 6456 \AA ~and 6728 \AA , 
and they generally have a S/N $\approx 340$ pixel$^{-1}$ in the continuum.
The spectra were extracted and calibrated using standard routines 
in {\it IRAF}\footnote{IRAF is distributed by the National Optical Astronomy 
Observatory, which is operated by the Association of Universities 
for Research in Astronomy, Inc., under cooperative agreement 
with the National Science Foundation.}, and then each continuum 
rectified spectrum was transformed onto a uniform heliocentric 
wavelength grid for analysis.  Atmospheric telluric lines were removed 
by division of modified versions of spectra of the rapidly rotating
A-star $\zeta$~Aql that we also observed \citep{gie02a}.

We measured radial velocities for HD~1383 using a template fitting 
method \citep{gie02b}.  The red spectrum of HD~1383 has few lines in the
region observed.  H$\alpha$ was not used in the radial velocity 
analysis because it is too broad and the components are generally blended.  
The lines of \ion{C}{2} $\lambda\lambda 6578, 6582$ and 
\ion{O}{2} $\lambda\lambda 6641, 6721$ are very weak and 
difficult to measure in individual spectra.
Thus, we focused on the remaining, problem-free 
line of \ion{He}{1} $\lambda6678$ for this radial velocity study.  
The individual component lines are quite narrow, symmetric, 
and comparable in shape to the instrumental broadening function, so we chose to 
represent each component's profile as a Gaussian function.  
We selected appropriate shape parameters by fitting 
Gaussians to the best separated profiles of \ion{He}{1} 
$\lambda 6678$, and we used the average values of the parameters 
to create separate Gaussian absorption line 
profiles to represent both the primary and secondary stars.
We also obtained preliminary orbital velocity curves based upon the 
Gaussian fits of the well separated spectra, and these were used 
to estimate the approximate velocities for all the times of observation.
We then determined radial velocities for both components in 
all of our spectra by a non-linear, least-squares fit of the composite 
profiles, and our results are collected in Table~1, which 
lists the heliocentric Julian date of mid-exposure, orbital phase
from the solution for the primary component, and the observed 
velocity plus the residual from the fit (observed minus 
calculated) for both components.

\placetable{tab1}      


\section{Orbital Elements of HD~1383}        

\citet{hil86} found that HD~1383 is a double-lined spectroscopic 
binary with a period of 20.2819~days.  We combined our radial 
velocities (Table~1) with the compilation of radial velocity 
measurements from \citet{hil86} for a total 101 and 77 radial 
velocity measurements (spanning 75 y) of the primary and secondary 
components, respectively.  Using a 'dirty' discrete Fourier
transform and CLEAN deconvolution algorithm \citep*{rob87}, 
we constructed power spectra for both the primary 
and secondary using the time series of radial velocity 
measurements for each.  The strongest signal in the power 
spectra occurs at $P = 20.3$ days, which we then used as a starting value 
for the following orbital solutions.

We used the non-linear, least-squares fitting program from \citet{mor74} 
to determine orbital elements.  We found that our derived velocities were 
swapped between the primary and secondary for two observations 
when the components were thoroughly blended, and we assigned 
zero weight to these measurements in our orbital fit.  
First, we determined the period of the 
primary and secondary independently using all the data available, 
and then we determined a mean value of $P = 20.28184 \pm 0.00020$ d, which we 
fixed for both stars in the subsequent orbital solutions for HD~1383.  
Then only the new radial velocity data presented in this paper were used to 
calculate independent orbital elements for the primary and secondary stars.  
The separate results for the primary and secondary are presented in 
Table~2 together with the original results from \citet{hil86}, 
and the radial velocity curves and observations are plotted in Figure~1.  
We find that the orbital elements for the primary 
and secondary are mainly consistent with each other and with the 
original determinations by \citet{hil86} with two interesting 
exceptions.  First, the star identified as the ``primary'' by 
\citet{hil86} turns out to be the lower mass object in our solution
because of a slight revision in the semiamplitudes.  Rather than
introducing more confusion about the stars' identities, we will retain 
the labels of primary and secondary given by \citet{hil86}.
Second, we find that the eccentricity derived for the primary 
is approximately $3\sigma$ different from that obtained for
the secondary.  Furthermore, the longitude of periastron values 
are suspiciously close to $0^\circ$ and $180^\circ$, which 
suggests that the velocity curves may be distorted by subtle 
emission effects from circumstellar gas (the possible origin of 
the non-uniform distribution of longitude of periastron among
massive binaries known as the Barr Effect; \citealt{bat68,fra79,how93}).
Given these difficulties, we decided not to force a joint solution 
with a common geometry and systemic velocity.

\placetable{tab2}      

\begin{figure}
\begin{center}
{\includegraphics[angle=90,height=12cm]{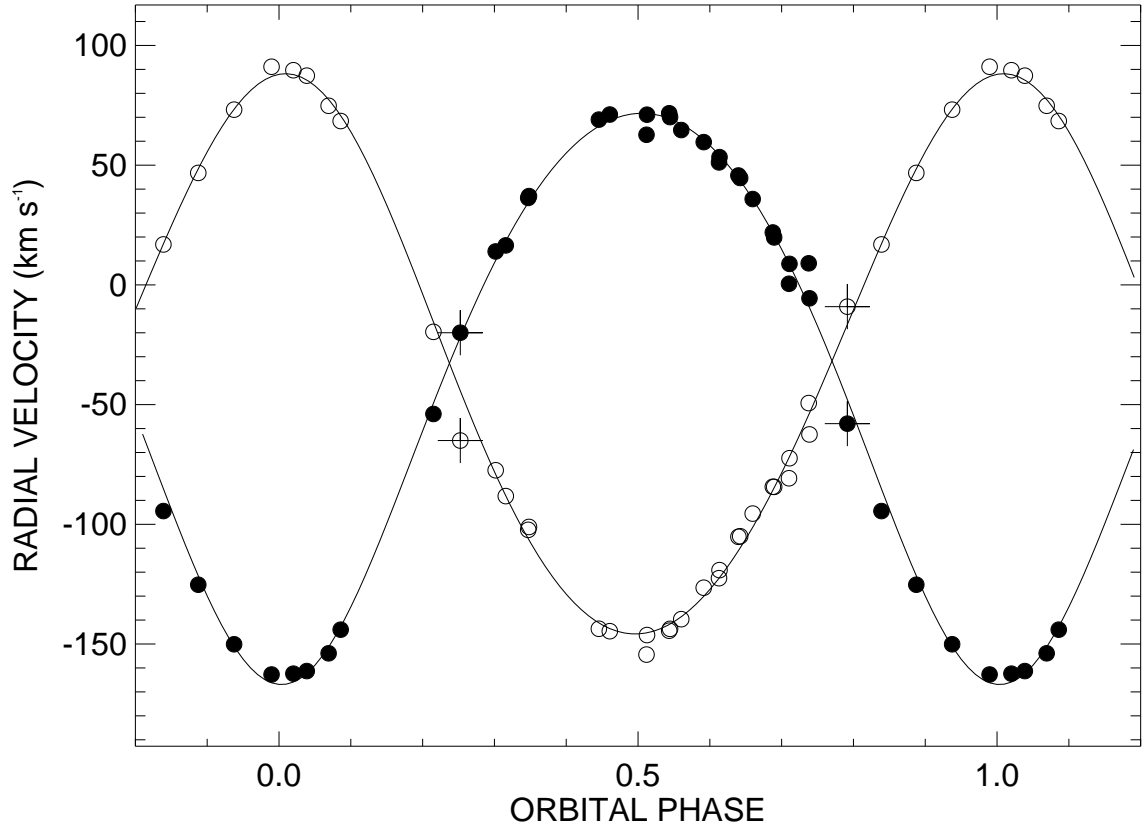}}
\end{center}
\caption{Calculated radial velocity curves ({\it solid lines}) for HD~1383.  
The errors in the measured radial velocities of the primary star ({\it filled circles}) 
and the secondary star ({\it open circles}) are comparable to the symbol sizes.  
Plus signs mark the measurements from two blended phases that were assigned zero weight.}
\label{fig1}
\end{figure}

\placefigure{fig1}     


\section{Tomographic Reconstruction and Stellar Parameters} 

Once the orbital solution was found for HD~1383, we used 
a tomographic reconstruction technique \citep{bag94} to separate the two 
individual spectra of the system.  The method of tomographic 
reconstruction uses all the combined spectra and 
their associated radial velocities to determine the 
appearance of each star's spectrum.  
The monochromatic flux ratio of the 
primary to the secondary was assumed to be 1.0
\citep{hil86}.  The ISM lines were removed from each spectrum prior to 
reconstruction to avoid spurious reconstructed features in their 
vicinity.  Figure~2 shows a plot of the reconstructed spectra 
with identifications of the principal lines. 
The two spectra are remarkably similar in the red spectral region.

\begin{figure}
\begin{center}
{\includegraphics[angle=90,height=12cm]{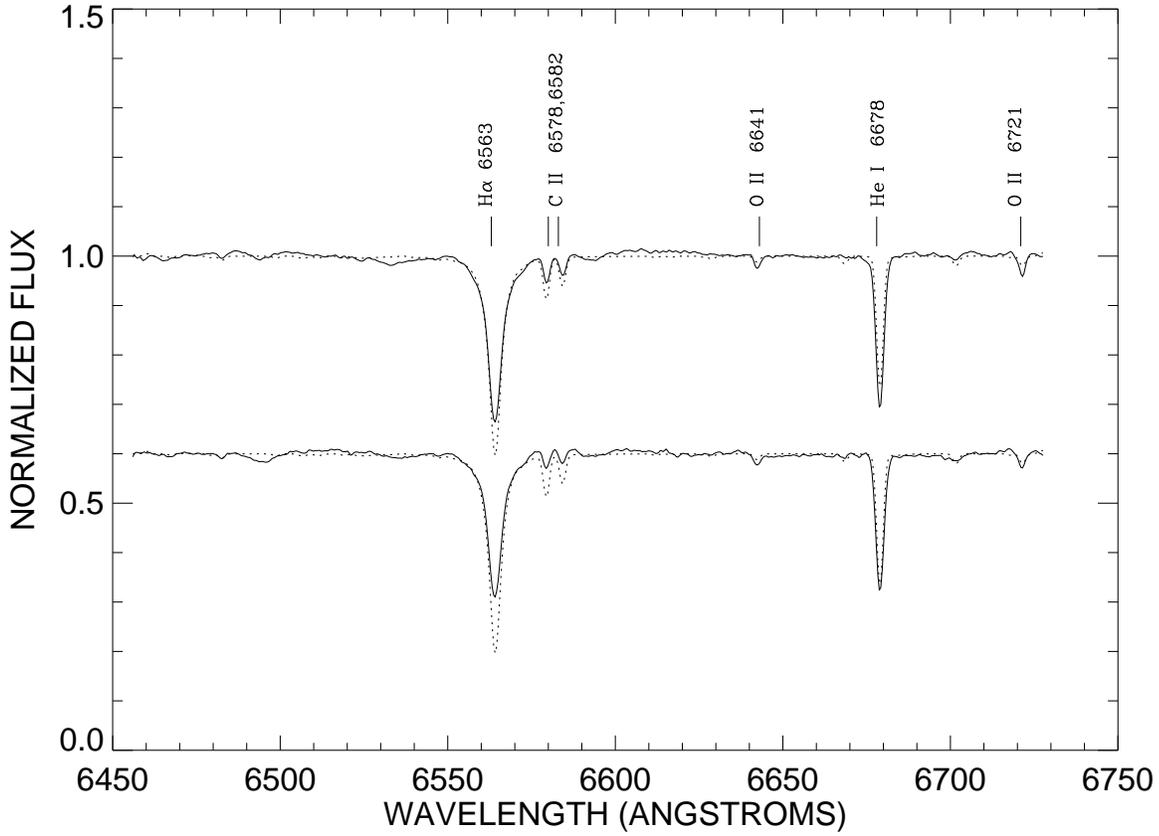}}
\end{center}
\caption{The tomographically reconstructed spectra ({\it solid lines}) of the 
(slightly more massive) secondary ({\it top}) and primary star ({\it bottom}) 
together with model spectra ({\it dotted lines}) for 
$T_{\rm eff} = 28000$~K, $\log g = 3.25$, and $V\sin i = 70$ \kms
~(a rotational broadening appropriate for \ion{He}{1} $\lambda 6678$).}
\label{fig2}
\end{figure}

\placefigure{fig2} 

We made estimates of the effective temperatures, gravities, and  
projected rotational velocities through a 
comparison with model spectra from the codes {\it TLUSTY} 
and {\it SYNSPEC} \citep*{hub88,hub95,hub98}.  \citet{lan03} 
presented a grid of model spectra for O-type stars that use line 
blanketed, non-local thermodynamic equilibrium, plane-parallel, 
hydrostatic atmospheres, and fortunately, these models extend to 
cool enough temperatures (27500~K) to be applicable to the 
the stars in HD~1383.  These models adopt a fixed microturbulent
velocity of 10 \kms, a value which is appropriate for B-supergiants
\citep{gie92}.  These models were also used by \citet{duf05} in 
their spectral analysis of B-supergiants in the SMC.

The projected rotational velocity, $V\sin i$, for each star 
was measured by comparing the observed FWHM of an absorption line
with that for model profiles for a range in assumed $V\sin i$.
The rotationally broadened profiles were calculated by 
a simple convolution of the zero-rotation model profiles 
with a rotational broadening function \citep{gra92} 
using a linear limb darkening coefficient $\epsilon=0.220$ 
(from the tabulated value for $T_{\rm eff} = 30000$~K, 
$\log g=3.0$, and $\lambda=6975$\AA ~from \citealt{wad85}).  
The derived projected rotational velocities based 
upon the \ion{He}{1} $\lambda 6678$ profile 
are $V\sin i = 76\pm6$ and $72\pm6$ \kms ~for 
the primary and secondary, respectively, in good agreement
with the estimate of $75\pm5$ \kms ~for both stars from 
measurements of blue spectral lines by \citet{hil86}. 
However, the weaker \ion{C}{2} $\lambda\lambda 6578, 6582$ 
lines are distinctly narrower and their mean projected rotational 
velocities are $V\sin i = 32\pm18$ and $26\pm18$ \kms 
~for the primary and secondary, respectively
(almost unresolved at our spectral resolution). 
\citet{rya02} argue that the line broadening of B-supergiants
is probably dominated by macroturbulence, so that the 
measured broadening only provides an upper limit on 
the actual rotational velocity.  They also find a trend 
for stronger lines (like \ion{He}{1} $\lambda 6678$) to 
display greater broadening than weaker lines 
(like \ion{C}{2} $\lambda\lambda 6578, 6582$), perhaps 
due to an increase in turbulent broadening with height in the
atmosphere.  Thus, the true projected rotational velocities of the 
components of HD~1383 are probably less than $\approx 30$ \kms. 

We then compared rotationally broadened versions of the model 
solar abundance spectra from \citet{lan03} directly with 
the reconstructed spectra to estimate temperatures and gravities.
The best matches were found with $T_{\rm eff}=28000\pm 1000$~K and 
$\log g=3.25\pm 0.25$ for both stars, and the model spectra 
for these parameters are shown in Figure~2 as dotted lines.  
This spectral region contains a number of features that 
are particularly sensitive to temperature and gravity. 
For example, at hotter temperatures the \ion{C}{2}, \ion{O}{2}, and \ion{He}{1} lines
weaken while new lines of \ion{Si}{4} $\lambda\lambda 6667, 6701$ and 
\ion{He}{2} $\lambda 6683$ appear that are clearly absent in the 
reconstructed spectra of HD~1383.  On the other hand, at cooler 
temperatures the \ion{C}{2} doublet increases greatly in strength and 
the \ion{N}{2} $\lambda6610$ line first appears (again absent in 
the reconstructed spectra).   The wings of the H$\alpha$ line 
provide a diagnostic of the gravity (wider due to greater linear  
Stark broadening in higher gravity models).  We caution that the 
core of H$\alpha$ appears to be filled in with residual emission 
from the stellar wind, and the TLUSTY models we used are based upon 
static atmospheres that do not account for wind outflow. 
However, we expect that the wind effects will be mainly confined 
to the higher opacity line core in relatively weak-wind stars like
those in HD~1383 and that the gravity derived from the 
pressure-broadened line wings will be close to (or slightly less)
than the actual gravity (see the discussion about the H$\gamma$ 
line wings in \citealt{pul96}).  The observed \ion{C}{2} lines 
appear to be somewhat weaker than predicted in the best matching model 
spectrum, which may reflect an underabundance of C caused by 
mixing of CNO-processed gas into the atmosphere.  
\citet*{mce99} also observed this effect in other B-supergiants.  
They give model fitting results for two galactic B0.5~Ib stars, HD~192422 
and HD~213087, and their derived temperatures and gravities are in 
reasonable agreement with our adopted values for HD~1383.  
We found that the generally good match between the model and observed 
spectra indicates that the monochromatic flux ratio is $1.0\pm 0.1$. 
 

\section{Discussion}                                

We can use our results to place some general constraints 
on the evolutionary status of the binary system.  These various
limits are summarized in a radius--mass diagram for the 
secondary star shown in Figure~3 (the corresponding 
diagram for the primary would appear almost the same). 
The system is not a known eclipsing binary, and we confirmed 
the lack of eclipses (or any other orbital phase-related variations) 
by plotting the available photometry 
from {\it Hipparcos} \citep{per97} as a function of orbital 
phase.  If we assume that the stars have the same radius $R$
(as indicated by their temperatures and the observed flux ratio),
then the upper limit on the orbital inclination $i$ set by 
the lack of eclipses is found from
\begin{equation}
\tan i = {{a\sin i} \over {2 R}} {{1 - e^{2}} \over {1 + e \cos \nu}}.
\end{equation}
We considered both conjunctions, $\nu = 90 \degr - \omega$ 
and $270 \degr - \omega_{2}$, with the derived eccentricities
for the primary and secondary to find the minimum inclination 
for a given radius and hence a lower limit on the mass of the 
secondary from $M_2\sin ^{3}i$ (Table~2),  
and the resulting radius--mass relationships are
plotted as dashed lines in Figure~3.     
The acceptable solution space is restricted to the region 
above these lines (at lower $i$).  
The next constraint comes from the gravity determination 
found by fitting the H$\alpha$ line wings, and the solid lines
in Figure~3 show the relations for $\log g = 3.25 \pm 0.25$. 
If weak stellar wind emission is biasing this measurement, 
then the actual $\log g$ may be somewhat larger than our estimate. 

\begin{figure}
\begin{center}
{\includegraphics[angle=90,height=12cm]{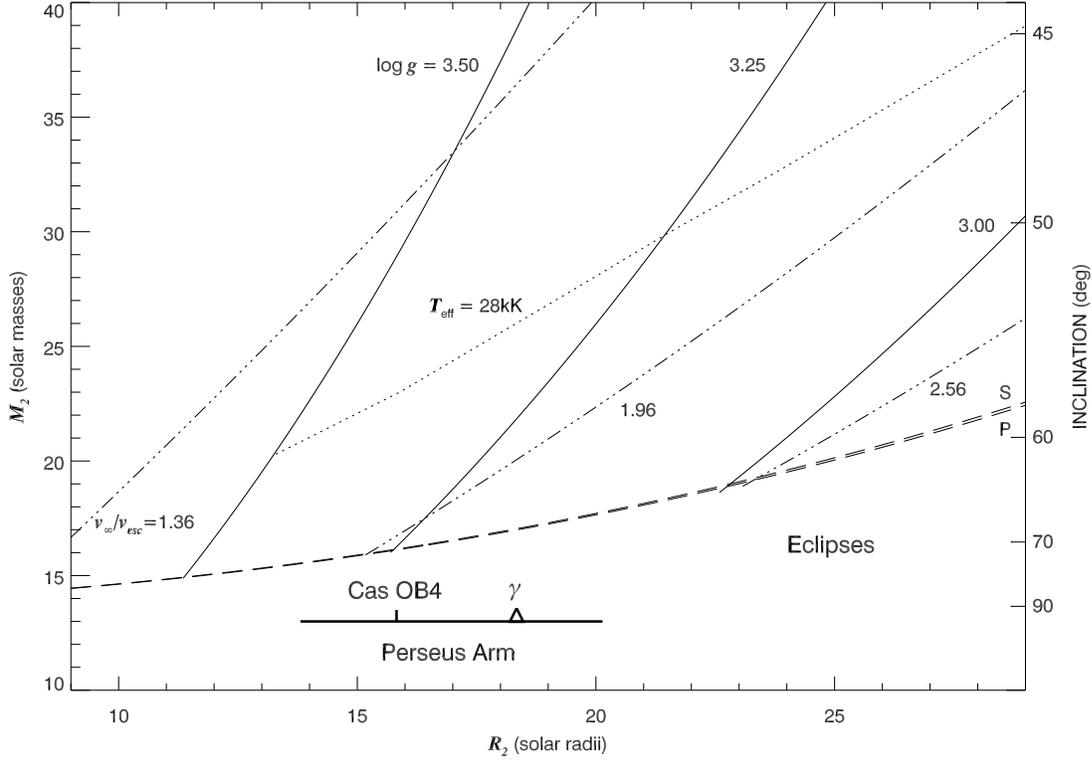}}
\end{center}
\caption{A plot of the possible range in secondary mass and radius.  
The system inclination for $M_2 \sin^{3}i$=13.7 is given on the right axis. 
The dashed lines mark the lower mass limit set by the absence of eclipses and
based upon the parameters from the primary ({\it P}) and secondary ({\it S})
orbital solutions.  The three solid lines show the relations for the 
range in gravity set by the H$\alpha$ line wings (indicated by values 
of $\log g$).  The three triple dot-dashed lines show the relations set by 
the span of values for the $v_\infty/v_{\rm esc}$ ratio (with the 
constant of proportionality labeled in each case). 
The dotted line indicates the single star evolutionary mass for 
a temperature of $T_{\rm eff} = 28000$~K \citep{sch92}. 
The bar at the bottom shows the corresponding radii for distances spanning a 
cut through the Perseus arm, the location of the Cas OB4 association, 
and the location where differential Galactic rotation matches the 
binary systemic velocity ($\gamma$).}
\label{fig3}
\end{figure}

\placefigure{fig3} 

We can use the stellar wind properties to find additional limitations. 
Theoretical and observational studies of the winds
of massive stars show that the wind terminal velocity $v_\infty$ is 
generally proportional to the escape velocity $v_{\rm esc}$ 
among stars of comparable temperature 
\citep*{pri90,lam95,kud00,eva04,cro06}. 
\citet{pri90} have made the most complete 
study of this relationship among the B-supergiants, and they 
find that $v_\infty = (1.96 \pm 0.60) v_{\rm esc}$ for B0-B3~I stars.  
We measured the terminal velocity
to be $v_\infty = 1100 \pm 120$ \kms ~according to the short wavelength 
absorption minimum point in the profile of the \ion{C}{4} $\lambda 1550$ 
P~Cygni line in a high dispersion spectrum of HD~1383 from the archive 
of the {\it International Ultraviolet Explorer Satellite} 
(made at orbital phase $\phi=0.21$, near conjunction).  
This terminal velocity is somewhat lower than the 
mean for the B0.5 supergiants of 1405 \kms ~but it is well within 
the range of terminal velocities for this group \citep{pri90}. 
We show the resulting radius--mass functions from the 
mean and $\pm 1\sigma$ limits of the $v_\infty / v_{\rm esc}$ relation
in Figure~3 ({\it triple dot-dashed line}).  Note that the larger 
values of this ratio found in recent studies \citep{eva04,cro06}
are probably more appropriate for much more luminous stars and 
would lead to unrealistically large radii in the case of HD~1383. 

Finally, we can obtain one more constraint by considering the 
radius--distance relationship that is established 
from fits of the reddened stellar flux distribution. 
We show in Figure~4 the observed spectral energy distribution for 
HD~1383 based upon low dispersion UV spectroscopy from {\it IUE},
Johnson $U, B, V$ magnitudes \citep*{hau70,col96}, and 2MASS $J, H, K$ infrared
magnitudes \citep*{coh03,cut03}.  We fit this flux distribution using a model 
spectrum from \citet{lan03} for two identical stars of $T_{\rm eff}=28000$~K
and $\log g = 3.25$, which we transformed using the Galactic extinction curve
from \citet{fit99}.  The best fit parameters for the extinction curve
are a reddening of $E(B-V)=0.55\pm0.05$ and a ratio of total-to-selective 
extinction of $R=2.97\pm0.15$.  The normalization of the model spectrum
yields the limb darkened angular diameter of one star, 
$\theta_{LD} = 54\pm 7$ microarcsec, and therefore the stellar radius is 
related to the distance $d$ (measured in kpc) by 
\begin{equation}
R/R_\odot = (5.8\pm 0.8) d.
\end{equation}

\begin{figure}
\begin{center}
{\includegraphics[angle=90,height=12cm]{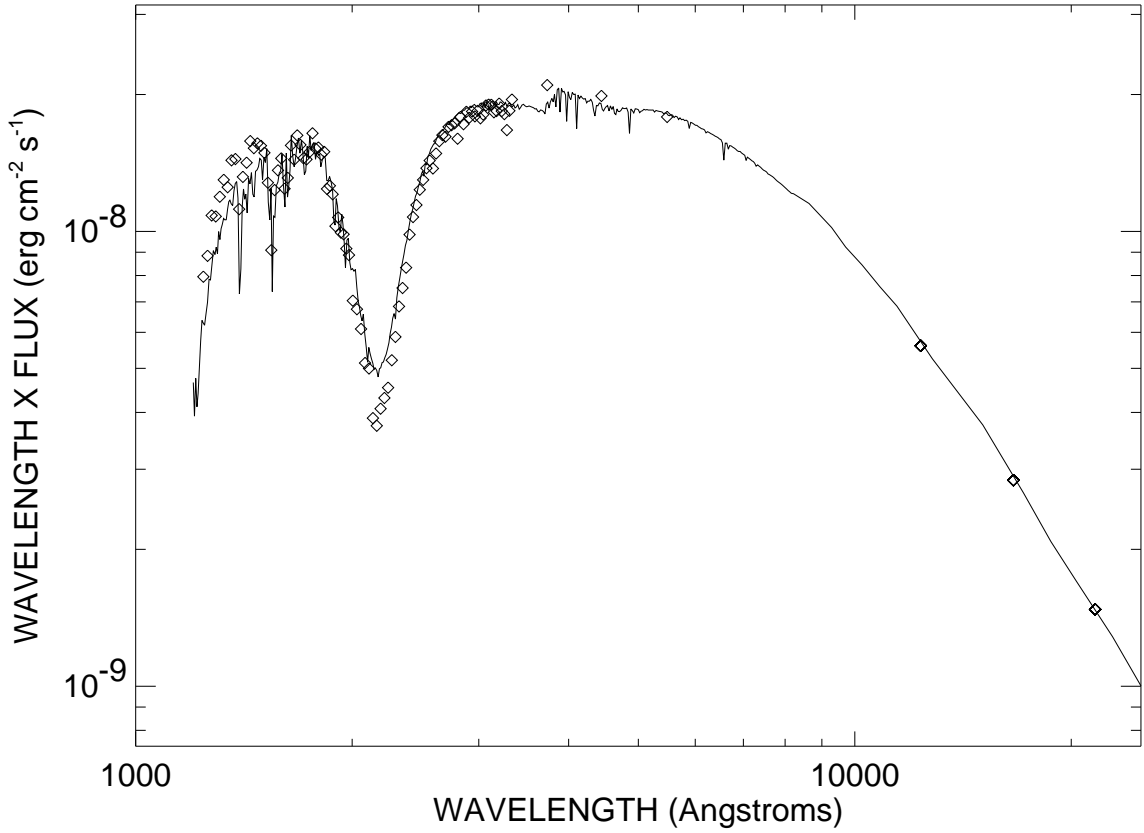}}
\end{center}
\caption{The spectral flux distribution and fit for the combined light of 
the HD~1383 components.  The fitting parameters are $T_{\rm eff} = 28000$~K, 
$\log g = 3.25$, $E(B-V)=0.55$ mag, $R=2.97$, and $\theta_{LD} = 54$ microarcsec for each star.}
\label{fig4}
\end{figure}

\placefigure{fig4} 

The binary is too distant to obtain a reliable parallax from 
{\it Hipparcos} measurements \citep{sch04}.
Our view through the plane of the Galaxy in the direction 
of HD~1383 ($l=119\fdg02$, $b=-0\fdg89$) traverses first the 
nearer Perseus arm ($d=2.4 - 3.5$~kpc) and then the more 
distant Cygnus arm ($d\gtrsim 3.9$~kpc) \citep{kim89,neg03}. 
We suspect that HD~1383 resides 
in the closer Perseus arm.  It is very close in the sky to 
BD$+60^\circ39$ (spectral classification of O9~V), which 
has an identical reddening and which \citet{gar92} assign 
to the Cas~OB4 association (at a distance of 2.8~kpc). 
There is a considerable amount of differential 
Galactic rotation along this line of sight, and we can 
estimate at what distance the systemic velocity of the 
binary matches the expected radial velocity difference
between the Sun and the remote local standard of rest. 
We used the procedure described by \citet{ber01} to 
find the distance--radial velocity relation along this
line of sight, and the binary's systemic velocity  
places it at a distance of 3.2~kpc (although if we allow a 
$\pm 10$ \kms ~deviation in motion from the local standard 
of rest, then the acceptable range is between 2.3 and 4.2~kpc). 
Both lines of evidence are consistent with a location in the 
Perseus arm, and we have plotted the corresponding stellar 
radius range as a solid line in the bottom of Figure~3. 

The combination of all these constraints indicates that
the parameter ranges are probably $R/R_\odot = 14 - 20$ and 
$M/M_\odot = 16 - 35$.   This range is consistent with the 
masses predicted by single star evolutionary tracks for the 
temperature and gravity of the components in HD~1383 
(illustrated in Figure~3 as a dotted line, from the evolutionary tracks 
for non-rotating, solar metallicity stars of $T_{\rm eff} = 28000$~K;
\citealt{sch92}).  The radii are much smaller than the Roche lobe radii
(approximately $44 R_\odot$ for masses of $25 M_\odot$), so 
the system is probably still observed in a pre-contact phase in which both 
stars have evolved like single objects. 

We originally selected HD~1383 as a possible target for exhibiting H$\alpha$ emission 
from colliding winds \citep{tha97}.  However, our spectra show no obvious signs 
of such H$\alpha$ emission.  We show in Figure~5 the H$\alpha$ 
profiles arranged as a function of orbital phase, and the 
variations appear to be entirely consistent with the motion 
of the photospheric H$\alpha$ lines of both components. 
\citet{san01} found evidence of a weak, broad, and stationary H$\alpha$ 
emission feature in their spectra of the similar colliding winds binary 
HD~152248, and they argue that the emission forms in a planar collision zone
between the stars.  The lack of H$\alpha$ colliding winds emission in 
the spectrum of HD~1383 is probably due to three significant differences 
between these binary systems.  First, the mass loss rates are lower
in HD~1383 compared to HD~152248.  We can estimate approximately 
the mass loss rates using the wind momentum relation for hot stars, 
\begin{equation}
\log (\dot{M} v_\infty (R/R_\odot)^{1/2}) = \log D_0 + x \log (L/L_\odot) 
\end{equation}
where $D_0$ and $x$ are constants and $L$ is the stellar luminosity \citep*{kud00,vin00}.
If we adopt $T_{\rm eff}=28000$~K, $R/R_\odot = 14 - 20$, and $v_\infty=1100$ \kms, 
then the predicted mass loss rate is $\log \dot{M} = -6.3 \pm 0.4$
(units of $M_\odot$~y$^{-1}$) according to the model of \citet{vin00}, 
which is about six times lower than the mass loss rates of the stars in 
HD~152248 \citep{san04}.  Second, the binary separation is about twice as 
large in HD~1383 compared to HD~152248 \citep{san04}, and hence the 
wind density in the regions near the collision zone will be much lower. 
Third, the collision zone itself may cool less efficiently and may avoid 
the formation of high density gas fragments that are predicted to occur 
according to hydrodyanmical simulations of the winds of HD~152248 \citep{san04}.  
\citet*{ste92} show how the gas dynamics of the colliding winds zone
depends on the ratio of the cooling timescale to the gas flow timescale, 
$\chi \approx v_8^4 ~d_{12} / \dot{M}_{-7}$, where $v_8$ is the wind 
velocity in units of 1000 km~s$^{-1}$, $d_{12}$ is the distance from 
the star to the contact surface in units of $10^7$ km, and 
$\dot{M}_{-7}$ is the mass loss rate in units of $10^{-7} ~M_\odot$ y$^{-1}$.
This ratio is about 1.1 for HD~1383, indicating that the collision zone 
remains hot over dimensions comparable to those of the binary system 
(an adiabatic wind zone), but the ratio is much smaller (0.1) in the 
case of HD~152248 where efficient cooling (in a radiative colliding wind) leads to  
the fragmentation of the shock front into knots of cool gas \citep{san04}. 
Since H$\alpha$ emission is a recombination process that depends on the 
square of the gas density and since the colliding wind density will be
much lower in HD~1383 compared to HD~152248 for all of the reasons outlined above, 
the apparent lack of H$\alpha$ emission in the spectrum of HD~1383 is not 
surprising. 
 
\begin{figure}
\begin{center}
{\includegraphics[angle=0,height=12cm]{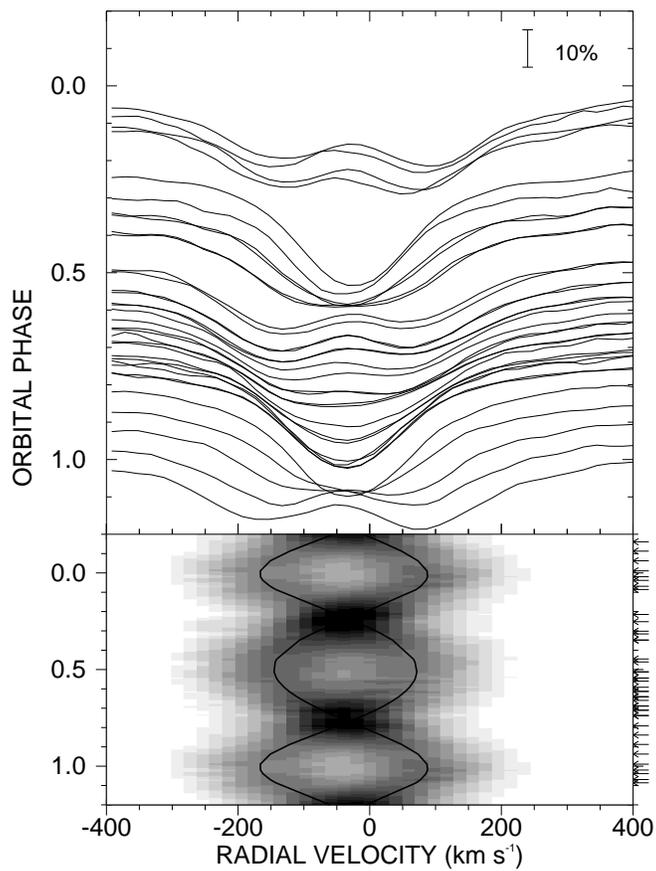}}
\end{center}
\caption{{\it Top:} H$\alpha$ line profiles plotted against 
heliocentric radial velocity.  The continuum of each observation is
normalized so that the $y$-ordinate equals the 
primary star phase at the time of observation.
{\it Bottom:}  A gray scale plot that shows the phase and velocity 
variations of the H$\alpha$ line profiles shown above.  Specific 
times of individual measurements are indicated by arrows on the 
right hand side. There are 16 gray levels that are determined by the difference
in intensities between the minimum to maximum observed values 
for all spectra.  The phase has been wrapped to enhance the 
sense of phase continuity, and the calculated radial velocity curves  
from the orbital solution are displayed as thick black lines.}
\label{fig5}
\end{figure}

\placefigure{fig5} 

Our results indicate that HD~1383 is a wide enough system that the 
components have avoided direct interaction or mass exchange. 
According to the models of \citet{sch92}, this state may last for 
another 0.5~Myr (for masses of $25 M_\odot$).  However, after that 
time both stars will quickly grow in radius and reach contact within 
the last $10^4$~y before they explode as supernovae.  Their brief 
interaction phase then may result in a common envelope stage 
leading to a shorter period system containing an O-supergiant 
and neutron star (like the massive X-ray binary system HD~153919/4U1700--37; 
\citealt{ank01}) or in a wider binary consisting of a
rapidly evolving B-A supergiant transferring mass at a tremendous rate
to a collapsed companion surrounded by a super-Eddington accretion disk 
(like SS~433; \citealt{hil04}).  Either way, HD~1383 is destined to 
become an extraordinarily energetic interacting binary for a brief 
instant in the Galaxy's future.


\acknowledgments

We thank Daryl Willmarth and the staff of KPNO for their assistance 
in making these observations possible.  We are also grateful 
to the referee, Dr.\ Hughes Sana, for his insight and helpful 
comments on this study.   This work was  
supported by the National Science Foundation under Grant No.~AST-0205297.
Institutional support has been provided from the GSU College
of Arts and Sciences and from the Research Program Enhancement
fund of the Board of Regents of the University System of Georgia,
administered through the GSU Office of the Vice President
for Research.  




\newpage
\begin{deluxetable}{lccccc}
\tabletypesize{\scriptsize}
\tablewidth{0pt}
\tablenum{1}
\tablecaption{HD 1383 Radial Velocity Measurements\label{tab1}}
\tablehead{

\colhead{HJD} &
\colhead{Primary} &
\colhead{$V_1$} &
\colhead{$(O-C)_1$}	&
\colhead{$V_2$}	&
\colhead{$(O-C)_2$}	\\
\colhead{($-$2,451,000)} &
\colhead{Orbital Phase\tablenotemark{a}} &
\colhead{(km s$^{-1}$)} &
\colhead{(km s$^{-1}$)}	&
\colhead{(km s$^{-1}$)} &
\colhead{(km s$^{-1}$)} }
\startdata
419.951\tablenotemark{b}\dotfill&0.253& \phn$-$19.9     &       \phn\phs1.9     &       \phn$-$65.0     &           $-$20.7     \\
420.950\dotfill	&	0.302	&	\phn\phs14.0	&	\phn\phs3.7	&	\phn$-$77.4	&	\phn\phs1.7	\\
421.869\dotfill	&	0.347	&	\phn\phs36.3	&	\phn\phs2.0	&	$-$102.3	&	\phn\phs3.7	\\
421.891\dotfill	&	0.348	&	\phn\phs37.1	&	\phn\phs2.3	&	$-$101.0	&	\phn\phs5.5	\\
423.865\dotfill	&	0.445	&	\phn\phs69.0	&	\phn\phs2.7	&	$-$143.6	&	\phn$-$2.4	\\
425.851\dotfill	&	0.543	&	\phn\phs71.7	&	\phn\phs2.2	&	$-$144.4	&	\phn$-$2.5	\\
425.875\dotfill	&	0.545	&	\phn\phs70.1	&	\phn\phs0.7	&	$-$143.6	&	\phn$-$2.0	\\
426.827\dotfill	&	0.591	&	\phn\phs59.6	&	\phn$-$0.9	&	$-$126.4	&	\phn\phs3.6	\\
427.806\dotfill	&	0.640	&	\phn\phs45.7	&	\phn\phs1.2	&	$-$105.2	&	\phn\phs5.8	\\
427.856\dotfill	&	0.642	&	\phn\phs44.6	&	\phn\phs1.1	&	$-$104.9	&	\phn\phs4.9	\\
428.778\dotfill	&	0.688	&	\phn\phs21.9	&	\phn\phs0.1	&	\phn$-$84.3	&	\phn\phs1.5	\\
428.816\dotfill	&	0.690	&	\phn\phs19.8	&	\phn$-$0.9	&	\phn$-$84.3	&	\phn\phs0.4	\\
429.792\dotfill	&	0.738	&	\phn\phn\phs9.0	&	\phs17.8	&	\phn$-$49.3	&	\phn\phs4.9	\\
429.813\dotfill	&	0.739	&	\phn\phn$-$5.6	&	\phn\phs3.9	&	\phn$-$62.4	&	\phn$-$8.9	\\
464.737\dotfill	&	0.461	&	\phn\phs71.1	&	\phn\phs2.5	&	$-$144.6	&	\phn$-$1.0	\\
465.774\dotfill	&	0.512	&	\phn\phs62.7	&	\phn$-$8.9	&	$-$154.4	&	\phn$-$9.0	\\
465.788\dotfill	&	0.512	&	\phn\phs71.1	&	\phn$-$0.5	&	$-$146.2	&	\phn$-$0.9	\\
466.756\dotfill	&	0.560	&	\phn\phs64.7	&	\phn$-$2.4	&	$-$139.6	&	\phn$-$1.0	\\
467.822\dotfill	&	0.613	&	\phn\phs51.1	&	\phn$-$3.2	&	$-$122.4	&	\phn\phs0.0	\\
467.836\dotfill	&	0.613	&	\phn\phs53.3	&	\phn$-$0.7	&	$-$119.1	&	\phn\phs3.1	\\
468.773\dotfill	&	0.660	&	\phn\phs35.9	&	\phn\phs0.0	&	\phn$-$95.5	&	\phn\phs5.8	\\
469.799\dotfill	&	0.710	&	\phn\phn\phs0.5	&	\phn$-$8.3	&	\phn$-$80.7	&	\phn$-$8.5	\\
469.813\dotfill	&	0.711	&	\phn\phn\phs8.8	&	\phn\phs0.4	&	\phn$-$72.3	&	\phn$-$0.6	\\
491.730\tablenotemark{b}\dotfill&0.792&	\phn$-$57.9	&	$-$10.0		&	\phn\phn$-$9.1	&	\phn\phs7.4	\\
492.693\dotfill	&	0.839	&	\phn$-$94.4	&	\phn$-$9.2	&	\phn\phs16.9	&	\phn$-$0.2	\\
493.677\dotfill	&	0.888	&	$-$125.2	&	\phn$-$3.6	&	\phn\phs46.8	&	\phn$-$1.7	\\
494.689\dotfill	&	0.937	&	$-$150.1	&	\phn\phs0.9	&	\phn\phs73.2	&	\phn$-$0.2	\\
495.747\dotfill	&	0.990	&	$-$162.7	&	\phn\phs3.5	&	\phn\phs91.1	&	\phn\phs4.1	\\
496.742\dotfill	&	0.039	&	$-$161.3	&	\phn\phs1.0	&	\phn\phs87.4	&	\phn\phs1.8	\\
497.700\dotfill	&	0.086	&	$-$144.0	&	\phn$-$1.2	&	\phn\phs68.4	&	\phn$-$2.2	\\
516.645\dotfill	&	0.020	&	$-$162.3	&	\phn\phs3.5	&	\phn\phs89.6	&	\phn\phs1.8	\\
517.641\dotfill	&	0.069	&	$-$153.8	&	\phn$-$2.6	&	\phn\phs74.8	&	\phn$-$2.6	\\
520.601\dotfill	&	0.215	&	\phn$-$53.9	&	\phn$-$4.4	&	\phn$-$19.6	&	\phn$-$4.1	\\
522.644\dotfill	&	0.316	&	\phn\phs16.5	&	\phn$-$1.8	&	\phn$-$88.2	&	\phn$-$0.1	\\
\enddata
\tablenotetext{a}{Secondary Phase = Primary Phase -- 0.051}
\tablenotetext{b}{Primary -- secondary swapped velocities are given and assigned zero weight.}
\end{deluxetable}

\newpage

\begin{deluxetable}{lcc}
\tabletypesize{\scriptsize}
\tablewidth{0pc}
\tablenum{2}
\tablecaption{Orbital Elements for HD 1383\label{tab2}}
\tablehead{
\colhead{Element}	& 
\colhead{\citet{hil86}}	&
\colhead{This Work}	}
\startdata
$P$~(days)                 \dotfill & 20.2819\tablenotemark{a}& 20.28184\tablenotemark{a}  \\
$T_1$ (HJD--2,400,000)     \dotfill & \nodata	              & $51414.8 \pm 0.4$\phn\phn\phn\phn \\
$T_2$ (HJD--2,400,000)     \dotfill & \nodata	              & $51415.9 \pm 0.6$\phn\phn\phn\phn \\
$e_1$                      \dotfill & $0.076\pm0.024$         & $0.116 \pm 0.012$	\\
$e_2$                      \dotfill & $0.027\pm0.028$         & $0.069 \pm 0.009$	\\
$\omega_1$ (deg)           \dotfill & $181 \pm 10$\phn        & $178 \pm 6$\phn\phn  \\
$\omega_2$ (deg)           \dotfill & $355 \pm 29$\phn        & $17 \pm 10$  \\
$K_1$ (km s$^{-1}$)        \dotfill & $113 \pm 1$\phn\phn & $119 \pm 1$\phn\phn  \\
$K_2$ (km s$^{-1}$)        \dotfill & $117 \pm 2$\phn\phn & $117 \pm 1$\phn\phn  \\
$\gamma_1$ (km s$^{-1}$)   \dotfill & $-35.1 \pm 1.8$\phs\phn & $-33.8 \pm 1.0$\phn\phs  \\
$\gamma_2$ (km s$^{-1}$)   \dotfill & $-34.7 \pm 2.1$\phs\phn & $-36.5 \pm 0.8$\phn\phs  \\
$q$ ($M_2/M_1$)            \dotfill & $0.968 \pm 0.018$       & $1.02\pm 0.01$ \\
$M_1\sin ^{3}i$ ($M_\odot$)\dotfill & $12.7 \pm 0.2$\phn      & $13.7 \pm 0.2$\phn  \\
$M_2\sin ^{3}i$ ($M_\odot$)\dotfill & $12.4 \pm 0.2$\phn      & $13.7 \pm 0.2$\phn  \\
$a \sin i$ ($R_\odot$)     \dotfill & $92.2 \pm 0.8$\phn      & $94.2 \pm 0.6$\phn  \\
$\sigma_1$ (km s$^{-1}$)   \dotfill & $8.0$                   & $5.1$  \\
$\sigma_2$ (km s$^{-1}$)   \dotfill & $9.6$                   & $4.3$  \\
\enddata
\tablenotetext{a}{Fixed.}
\end{deluxetable}



\clearpage

\input{epsf}

\end{document}